\documentclass{iau}
\usepackage{graphicx}

\title[Bisector analysis of RR Lyrae] 
{Bisector analysis of RR Lyrae: atmosphere dynamics at different phases}

\author[E. Guggenberger et al.]   
{Elisabeth Guggenberger$^1$, Denis Shulyak$^2$, Vadim Tsymbal$^3$ and Katrien Kolenberg$^{4,5}$
}

\affiliation{$^1$Institut f\"ur Astrophysik, Universit\"at Wien, T\"urkenschanzstrasse 17,
1180 Vienna, Austria\\ email: {\tt elisabeth.guggenberger@univie.ac.at} \\[\affilskip]
$^2$ Institute of Astrophysics, Georg-August University, Friedrich-Hund-Platz 1, 37077, G\"ottingen, Germany \\[\affilskip]
$^3$ Tavrian National University, VernadskiyÕs Avenue 4, Simferopol, Crimea, 95007, Ukraine \\[\affilskip]
$^4$ Harvard-Smithsonian Center for Astrophysics, 60 Garden Street, Cambridge, MA 02138  \\[\affilskip]
$^5$ Instituut voor Sterrenkunde, University of Leuven, Celestijnenlaan 200D, B 3001 Heverlee, Belgium \\[\affilskip]}

\pubyear{2013}
\volume{301}  
\pagerange{}
\jname{Precision Asteroseismology}
\editors{}
\begin{document}

\maketitle

\begin{abstract}
This article reports some preliminary results on an analysis of line bisectors of metal absorption lines of
RR Lyrae, the prototype of its class of pulsators. The extensive data set
used for this study consists of a time series of spectra obtained at
various pulsation phases as well as different Blazhko phases. This setup should allow a comparison of the atmospheric behavior, especially of the function of radial velocity versus depth at differing Blazhko phases, but
(almost) identical pulsation phase, making it possible to investigate
whether the modulation causes a change in the atmospheric motion of RR
Lyrae.
While the nature of the Blazhko modulation has often been investigated
photometrically and described as a change in the light curve, studies on
time series of high resolution spectra are rare. We present for the first
time work on line bisectors at different phases of RR Lyr.
\keywords{line: profiles, techniques: spectroscopic, stars: atmospheres, stars: variables: RR Lyrae}
\end{abstract}

\firstsection 
\section{Introduction}

Dedicated photometric surveys such as for example the Konkoly survey (\cite{jur09}) as well as high-precision satellite data for example from the \textit{Kepler} mission (\cite{benko}) have recently been very successful at revealing many detailed characteristics of RR Lyrae stars and their major unsolved problem: the Blazhko effect. Both from modeling and from spectroscopic studies it is known, however, that violent dynamic phenomena happen during the pulsation in the atmosphere of RR Lyrae stars. An important question is how the motion of the atmosphere changes when the Blazhko phase changes. As the pulsational motion of the atmosphere can best be studied spectroscopically, especially with the help of line asymmetries, we have chosen a method for our study that has never before been applied to RR Lyrae stars: a bisector analysis.

Spectral line bisectors (which are defined as the midpoint of a horizontal line connecting the two sides of the line profile) are a powerful tool to quantitatively study line asymmetries, allowing to extract information about velocity fields in the stellar atmosphere. The most well-known application of line bisectors is probably the study of solar and stellar granulation and convection (see for example \cite[Gray 2005]{gray2005}). 
A very different application of the analysis of bisectors is the detection of false positives in the hunt for extrasolar planets, where a change of the bisector during the observed radial velocity variation indicates that the signature in the radial velocity most likely does not originate from a planet but from stellar activity.

Another application of spectral line bisectors -- the one used in this study -- is the derivation of pulsational velocity fields, which relies on the fact that different parts of the line profile are formed at different depths of the stellar atmosphere. Using model atmospheres and synthetic spectra, it is possible to associate an observed part of the line profile with a specific layer in the star, allowing to derive atmospheric dynamics from the velocity information contained in the bisectors.

\begin{figure}[ht]
 \vspace*{-0.2 cm}
\begin{center}
 \includegraphics[width=13.5cm]{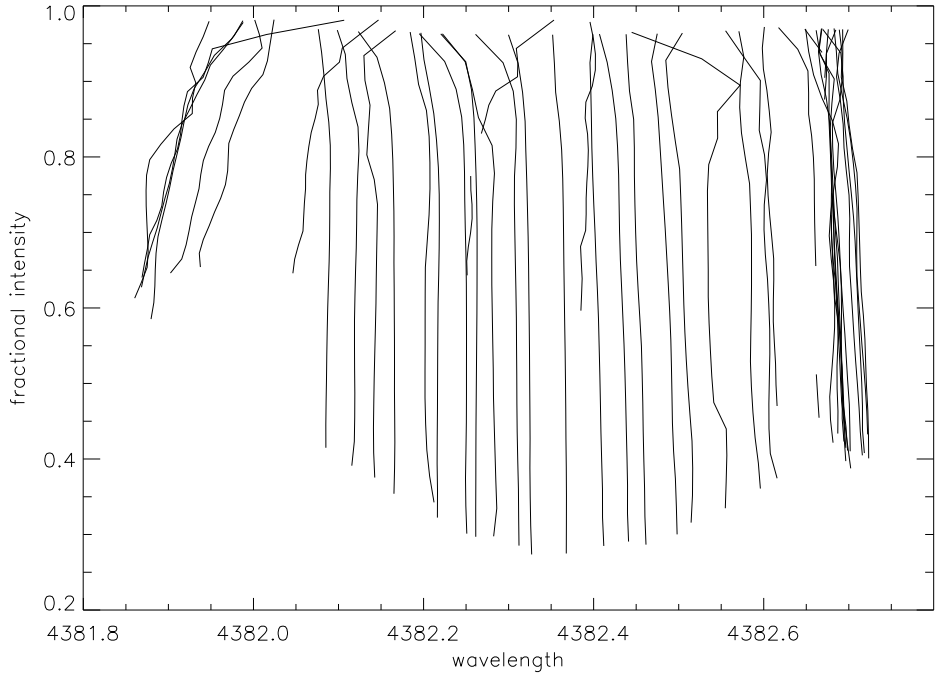} 
 \vspace*{-0.6 cm}
 \caption{The motion of a line bisector (the Fe~2 line at 4383.545 \AA~is shown as an example) during the pulsation. Motion is counterclockwise. The phase of maximum radius is the more or less straight bisector line in the middle of the plot when the line is very deep. Starting  from there, the star contracts, resulting in the redshifted lines in the right half of the Figure. Some tilted bisectors occur during the contraction. When the minimum radius is reached, lines become shallow, and distorted and tilted bisectors occur during expansion.}
   \label{bisectormotion}
\end{center}
\end{figure}

\section{The spectra, the models and the analysis}

The data set used for this analysis consists of 55 spectra with a resolution of R=60000 observed with the Robert G. Tull Coud\'e Spectrograph on the 2.7-m telescope of McDonald Observatory. 45 spectra were obtained near a Blazhko phase of 0.3 covering a full pulsation cycle, and 10 spectra were taken around a Blazhko phase of 0.8. A first analysis of these spectra as well as the technical details were presented by \cite{kol10} who selected the spectrum obtained at the most quiescent phase (near maximum radius) to perform an abundance analysis. During their analysis they also found that the microturbulence velocity ($v_{mic}$) is depth-dependent. The fundamental parameters, and specifically the element abundances, obtained by them for the ``quiet" phase serve as the basis for any future analysis of the remaining spectra.

In a second step, all the other spectra were subjected to a self-consistent detailed analysis. Fundamental parameters were determined for each spectrum, and the function of the depth-dependent $v_{mic}$ was found for each phase in an iterative process. The results of this extensive analysis will soon be published \cite{fossati13}. The \textsc{LLmodels} code (\cite{shulyak04}) was used to calculate static model atmospheres.

\begin{figure}[ht]
\begin{center}
 \includegraphics[width=13.5cm]{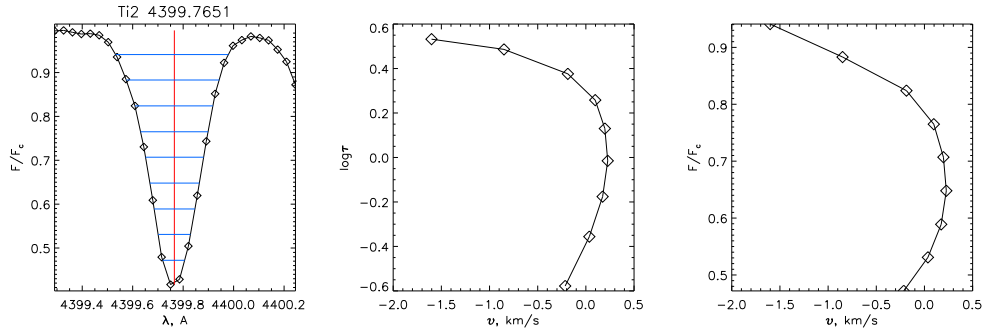} 
 \caption{Example of the bisectors derived for a  Ti~2 line. The left panel shows the observed spectral line (points) as well as the interpolation that was done to obtain the bisector points (line). Horizontal lines indicate the intensity values where bisector points were calculated. The middle panel shows the bisector with the optical depth on the y-axis, and the right panels shows the classical bisector as fractional intensity versus velocity.}
   \label{Ti2}
\end{center}
\end{figure}

For the bisector analysis, we selected all Fe~1, Fe~2, Ti~2, Cr~2, Ca~2 and Mg~1 lines that were unblended and therefore suitable for a reliable analysis, resulting in a sample of about 100 lines in total. Synthetic spectra were then calculated for all phases and for the complete wavelength range based on the final models obtained by \cite{fossati13}, using a new version of the \textsc{SynthV} code by Vadim Tsymbal (\cite{tsymbal96}). The bisector analysis code then extracts the ``classical" bisector (i.e., intensity versus velocity) as well as the bisector with formation depth (i.e. log($\tau$) versus velocity) for the selected lines and the selected spectra. 
The resulting motion of a single line bisector during the pulsation can be seen in Figure~\ref{bisectormotion} where an Fe~2 line is shown as an example. It is obvious that strong tilts and distortions of the bisector take place at some phases, and the wavelength shifts due to the strong pulsational motions can be seen. Figure~\ref{Ti2} gives a zoom-in revealing the curvature of a bisector of a Ti~2 line.\\
In the subsequent steps, a mean bisector will be calculated for each spectrum including all suitable lines to improve the accuracy, and a comparison to the predictions of pulsation models will be made. 

\section{Velocity curves and the Van Hoof effect}
\begin{figure}[ht]
 \vspace*{-0.6 cm}
\begin{center}
 \includegraphics[width=13.5cm]{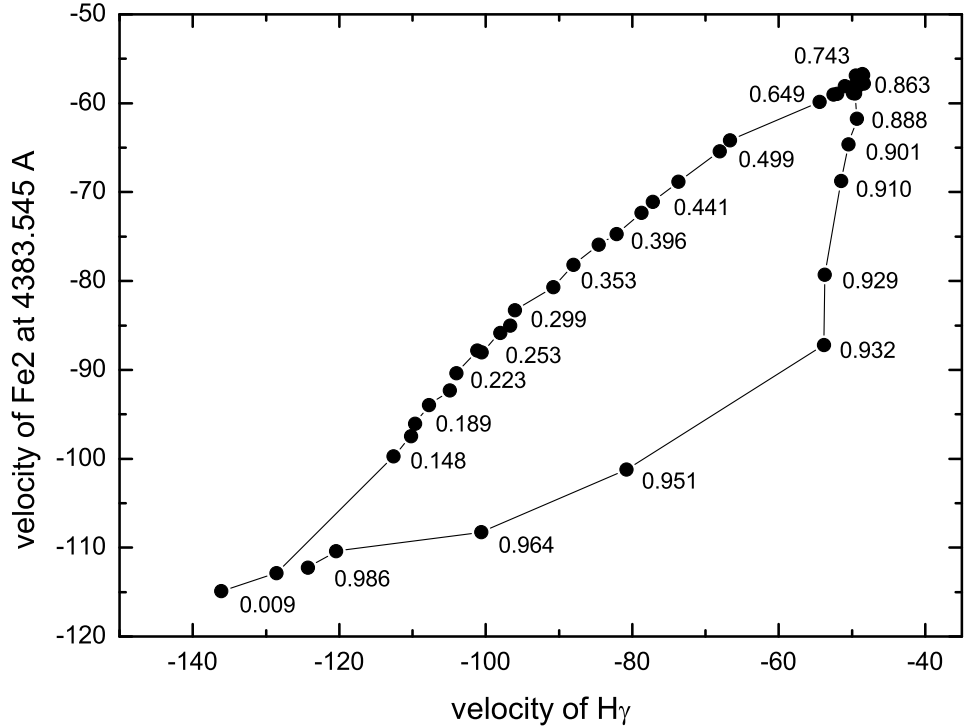} 
 \caption{A phase lag (Van Hoof effect) between the H$\gamma$ line and an Fe~2 line. The big area enclosed by the curve indicates a strong phase delay between a hydrogen line at the metal line, consistent with the running wave crossing the different line forming regions in the stellar atmosphere. Numbers in the plot indicate the pulsation phase.}
   \label{vanhoofpic}
\end{center}
 \vspace*{-0.2 cm}
\end{figure}

As a side result of our analysis we also obtain accurate velocity curves for every line that has been selected for the bisector analysis. These allow to study the Van Hoof effect, a phase lag between different layers in a pulsations star which has originally been diecovered in $\beta$ Cephei stars by \cite{vanhoof} and which has been shown to also exist in RR Lyrae stars by \cite{mathias95}. Following the method proposed by \cite{mathias93} who used the Van Hoof effect as a tool to study wave propagation in the atmosphere, we directly plot velocities of different lines against each other. As an example, the velocity of the H$\gamma$ line versus the velocity of a metal line is shown in Figure~\ref{vanhoofpic}. It can be seen that our results so far agree very well with the findings of \cite{mathias93}. As we have a huge number of unblended lines available and will automatically produce velocity curves for each of them, we expect that we can study this effect in some more detail. So far we have used the center of gravity to find the velocity of the line, but other methods will also be tested later in this project. 

\acknowledgements{
The author acknowledges support from the Austrian Science Fund (FWF), project number P19962-N16}. KK acknowledges support from Marie Curie Fellowship 255267 SAS-RRL within the 7th European Community Framework Program.

\end{document}